\documentclass[iop, twocolumn]{aastex6}
\hypersetup{citecolor=blue,urlcolor=red}
\usepackage{amsmath}
\usepackage{url}
\usepackage{enumerate}
\usepackage{multirow}
\usepackage{threeparttable}

\begin{document}
\title{Radio detection of an elusive millisecond pulsar in the Globular Cluster NGC~6397}

\author{
Lei Zhang$^{1*,2}$\href{https://orcid.org/0000-0001-8539-4237}{\includegraphics[scale=0.08]{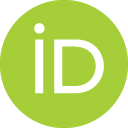}},
Alessandro Ridolfi$^{3,5}$\href{https://orcid.org/0000-0001-6762-2638}{\includegraphics[scale=0.08]{ORCIDiD.png}}, 
Harsha Blumer$^{4}$\href{https://orcid.org/0000-0003-4046-884X}{\includegraphics[scale=0.08]{ORCIDiD.png}}, 
Paulo C. C. Freire$^{5}$\href{https://orcid.org/0000-0003-1307-9435}{\includegraphics[scale=0.08]{ORCIDiD.png}},
Richard N. Manchester$^{6}$\href{https://orcid.org/0000-0001-9445-5732}{\includegraphics[scale=0.08]{ORCIDiD.png}}, 
Maura McLaughlin$^{4}$, 
Kyle Kremer$^{7.8}$\href{https://orcid.org/0000-0002-4086-3180}{\includegraphics[scale=0.08]{ORCIDiD.png}},
Andrew D. Cameron$^{2,9}$\href{https://orcid.org/0000-0002-2037-4216}{\includegraphics[scale=0.08]{ORCIDiD.png}},
Zhiyu Zhang$^{10,11}$\href{https://orcid.org/0000-0002-7299-2876}{\includegraphics[scale=0.08]{ORCIDiD.png}},  
Jan Behrend$^{5}$\href{https://orcid.org/0000-0002-8622-1298}{\includegraphics[scale=0.08]{ORCIDiD.png}},
Marta Burgay$^{3}$\href{https://orcid.org/0000-0002-8265-4344}{\includegraphics[scale=0.08]{ORCIDiD.png}}, 
Sarah Buchner$^{12}$\href{https://orcid.org/0000-0002-1691-0215}{\includegraphics[scale=0.08]{ORCIDiD.png}}, 
David J. Champion$^{5}$\href{https://orcid.org/0000-0003-1361-7723}{\includegraphics[scale=0.08]{ORCIDiD.png}}, 
Weiwei Chen$^{5}$\href{https://orcid.org/0000-0002-6089-7943}{\includegraphics[scale=0.08]{ORCIDiD.png}}, 
Shi Dai$^{13,1}$\href{https://orcid.org/0000-0002-9618-2499}{\includegraphics[scale=0.08]{ORCIDiD.png}}, 
Yi Feng$^{14}$\href{https://orcid.org/0000-0002-0475-7479}{\includegraphics[scale=0.08]{ORCIDiD.png}},
Xiaoting Fu$^{15}$\href{https://orcid.org/0000-0002-6506-1985}{\includegraphics[scale=0.08]{ORCIDiD.png}},    
Meng Guo$^{16,17}$, 
George Hobbs$^{6}$\href{https://orcid.org/0000-0003-1502-100X}{\includegraphics[scale=0.08]{ORCIDiD.png}}, 
Evan F. Keane$^{18}$\href{https://orcid.org/0000-0002-4553-655X}{\includegraphics[scale=0.08]{ORCIDiD.png}}, 
Michael Kramer$^{5}$\href{https://orcid.org/0000-0002-4175-2271}{\includegraphics[scale=0.08]{ORCIDiD.png}}, 
Lina Levin$^{19}$\href{https://orcid.org/0000-0002-2034-2986}{\includegraphics[scale=0.08]{ORCIDiD.png}}, 
Xiangdong Li$^{10,11}$, 
Mengmeng Ni$^{17}$,
Jingshan Pan$^{16,17}$,
Prajwal V. Padmanabh$^{5,20}$\href{https://orcid.org/0000-0001-5624-4635}{\includegraphics[scale=0.08]{ORCIDiD.png}}, 
Andrea Possenti$^{3,21}$\href{https://orcid.org/0000-0001-5902-3731}{\includegraphics[scale=0.08]{ORCIDiD.png}}, 
Scott M. Ransom$^{22}$,
Chao-Wei Tsai$^{1}$,\href{https://orcid.org/0000-0002-9390-9672}{\includegraphics[scale=0.08]{ORCIDiD.png}}, 
Vivek Venkatraman Krishnan$^{5}$\href{https://orcid.org/0000-0001-9518-9819}{\includegraphics[scale=0.08]{ORCIDiD.png}}, 
Pei Wang$^{1}$\href{https://orcid.org/0000-0002-3386-7159}{\includegraphics[scale=0.08]{ORCIDiD.png}},
Jie Zhang$^{23}$, 
Qijun Zhi$^{24}$,
Yongkun Zhang$^{1}$\href{https://orcid.org/0000-0002-8744-3546}{\includegraphics[scale=0.08]{ORCIDiD.png}},
Di Li$^{1*}$\href{https://orcid.org/0000-0003-3010-7661}{\includegraphics[scale=0.08]{ORCIDiD.png}}
}

\affiliation{
$^{1}$ {National Astronomical Observatories, Chinese Academy of Sciences, A20 Datun Road, Chaoyang District, Beijing 100101, China}\\
\textcolor{blue}{leizhang996@nao.cas.cn; dili@nao.cas.cn}\\
$^{2}$ {Centre for Astrophysics and Supercomputing, Swinburne University of Technology, P.O. Box 218, Hawthorn, VIC 3122, Australia}\\
$^{3}$ {INAF – Osservatorio Astronomico di Cagliari, Via della Scienza 5, I-09047 Selargius (CA), Italy}\\
$^{4}$ {Department of Physics and Astronomy, West Virginia University, Morgantown, WV 26506, USA}\\
$^{5}$ {Max-Planck Institut f{\"u}r Radioastronomie, Auf dem H{\"u}gel 69, D-53121 Bonn, Germany}\\
$^{6}$ {CSIRO Space and Astronomy, Australia Telescope National Facility, PO Box 76, Epping, NSW 1710, Australia}\\
$^{7}$ {TAPIR, California Institute of Technology, Pasadena, CA 91125, USA}\\
$^{8}$ {The Observatories of the Carnegie Institution for Science, Pasadena, CA 91101, USA}\\
$^{9}$ {ARC Center of Excellence for Gravitational Wave Discovery (OzGrav), Swinburne University of Technology, Mail H74, PO Box 218, VIC 3122, Australia}\\
$^{10}$ {School of Astronomy and Space Science, Nanjing University, Nanjing 210023, China}\\
$^{11}$ {Key Laboratory of Modern Astronomy and Astrophysics (Nanjing University), Ministry of Education, Nanjing 210093, People’s Republic of China}\\
$^{12}$ {South African Radio Astronomy Observatory, Fir Street, Black River Park, Cape Town, 7925, South Africa}\\
$^{13}$ {School of Science, Western Sydney University, Locked Bag 1797, Penrith, NSW 2751, Australia}\\
$^{14}$ {Research Center for Intelligent Computing Platforms, Zhejiang Laboratory, Hangzhou 311100, China}\\
$^{15}$ {Purple Mountain Observatory, Chinese Academy of Sciences, Nanjing 210023, China}\\
$^{16}$ {National Supercomputing Center in Jinan, Qilu University of Technology, 28666 East Jingshi Road, Licheng Distrist, Jinan 250103, China}\\
$^{17}$ {Jinan Institute of Supercomputing Technology, 28666 East Jingshi Road, Licheng Distrist, Jinan 250103, China}\\
$^{18}$ {School of Physics, Trinity College Dublin, College Green, Dublin 2, Ireland}\\
$^{19}$ {Jodrell Bank Centre for Astrophysics, Department of Physics and Astronomy, The University of Manchester, Manchester M13 9PL, UK}\\
$^{20}$ {Max-Planck-Institut f{\"u}r Gravitationsphysik (Albert-Einstein-Institut), D-30167 Hannover, Germany}\\
$^{21}$ {Dipartimento di Fisica, Universit$\acute{a}$ di Cagliari, S.P. Monserrato-Sestu Km 0,700, I-09042 Monserrato (CA), Italy}\\
$^{22}$ {NRAO, 520 Edgemont Road, Charlottesville, VA 22903, USA}\\
$^{23}$ {College of Physics and Electronic Engineering, Qilu Normal University, 2 Wenbo Road, Zhangqiu District, Jinan 250200, China}\\
$^{24}$ {Guizhou Provincial Key Laboratory of Radio Astronomy and Data Processing, Guizhou Normal University, Guiyang 550001, China}
}

\begin{abstract}
We report the discovery of a new 5.78 ms-period millisecond pulsar (MSP), PSR~J1740$-$5340B (NGC 6397B), in an eclipsing binary system discovered with the Parkes radio telescope (now also known as Murriyang\footnote{\url{https://blog.csiro.au/parkes-telescope-indigenous-name/}}), Australia, and confirmed with the MeerKAT radio telescope in South Africa. The measured orbital period, 1.97\,days, is the longest among all eclipsing binaries in globular clusters (GCs) and consistent with that of the coincident X-ray source U18, previously suggested to be a `hidden MSP'. Our \textit{XMM-Newton}  observations during NGC 6397B’s radio quiescent epochs detected no X-ray flares. NGC 6397B is either a transitional MSP or an eclipsing binary in its initial stage of mass transfer after the companion star left the main sequence. The discovery of NGC 6397B potentially reveals a subgroup of extremely faint and heavily obscured binary pulsars, thus providing a plausible explanation to the apparent dearth of binary neutron stars in core-collapsed GCs as well as a critical constraint on the evolution of GCs.
\end{abstract}

\section{Introduction}
Millisecond pulsars (MSPs) are evolved neutron stars (NSs) that have been spun up to short spin periods after a $\sim$Gyr-long phase of mass transfer in a low-mass X-ray binary (LMXB) phase~\citep{Alpar82}.
Globular clusters (GCs) are prolific environments for the formation of MSPs: their ultra-dense cores favor gravitational interactions between stars, which in turn result in the formation of various types of binary systems suitable for spinning up NSs. One subpopulation are known as ``redback" MSPs, which have hot and bloated main sequence-like $\gtrsim 0.1 M_{\odot}$ companions, and typically display irregular radio eclipses~\citep{Roberts13,Strader19}.

Core-collapsed GCs with steep inner density profiles, are fertile hunting grounds for exotic compact binary systems, containing white dwarfs (WDs), NSs and black holes (BHs). NGC 6397 is the closest ($\sim$2.3\,kpc) core-collapsed GC known~\citep{Marcano21}, possesses a central dark component with up to 2\% of its total mass~\citep{Vitral21}, likely composed primarily of WDs and NSs~\citep{Kremer21,Vitral22}. Past surveys, however, only confirmed one pulsar in this GC. Deep Chandra observations of the cluster revealed U18~\citep{Bogdanov10}, an X-ray source bearing similarity to the first confirmed redback radio MSP, PSR~J1740$-$5340A (or NGC~6397A, \citealt{DAmico01}, which also has a dominant non-thermal component, presumably from an intrabinary shock. A steep-spectrum pulsar-like radio counterpart was recently identified in ATCA images, further suggesting U18 to be a hidden MSP ~\citep{Zhao20}.  Further optical and UV observations with the Hubble Space Telescope suggested U18 to be a member of a redback system, with an orbital period of $1.96\pm0.06$ days~\citep{Marcano21}.

Using the ultra-wide-bandwidth low frequency receiver (UWL, \citealt{Hobbs20}) on the Parkes telescope, we have carried out radio observations of the globular cluster NGC 6397 to search for radio pulsations from the candidate MSP source U18. These were complemented by additional multi-wavelength observations made at other telescopes.
In Section 2, we describe the observations and data reduction. We present the results in Section 3 and discuss their implications in Section 4. We summarize our findings and provide our conclusions in Section 5.

\begin{figure*}
\begin{center}
  \includegraphics[angle=0,height=0.6\textwidth,width=0.9\textwidth]{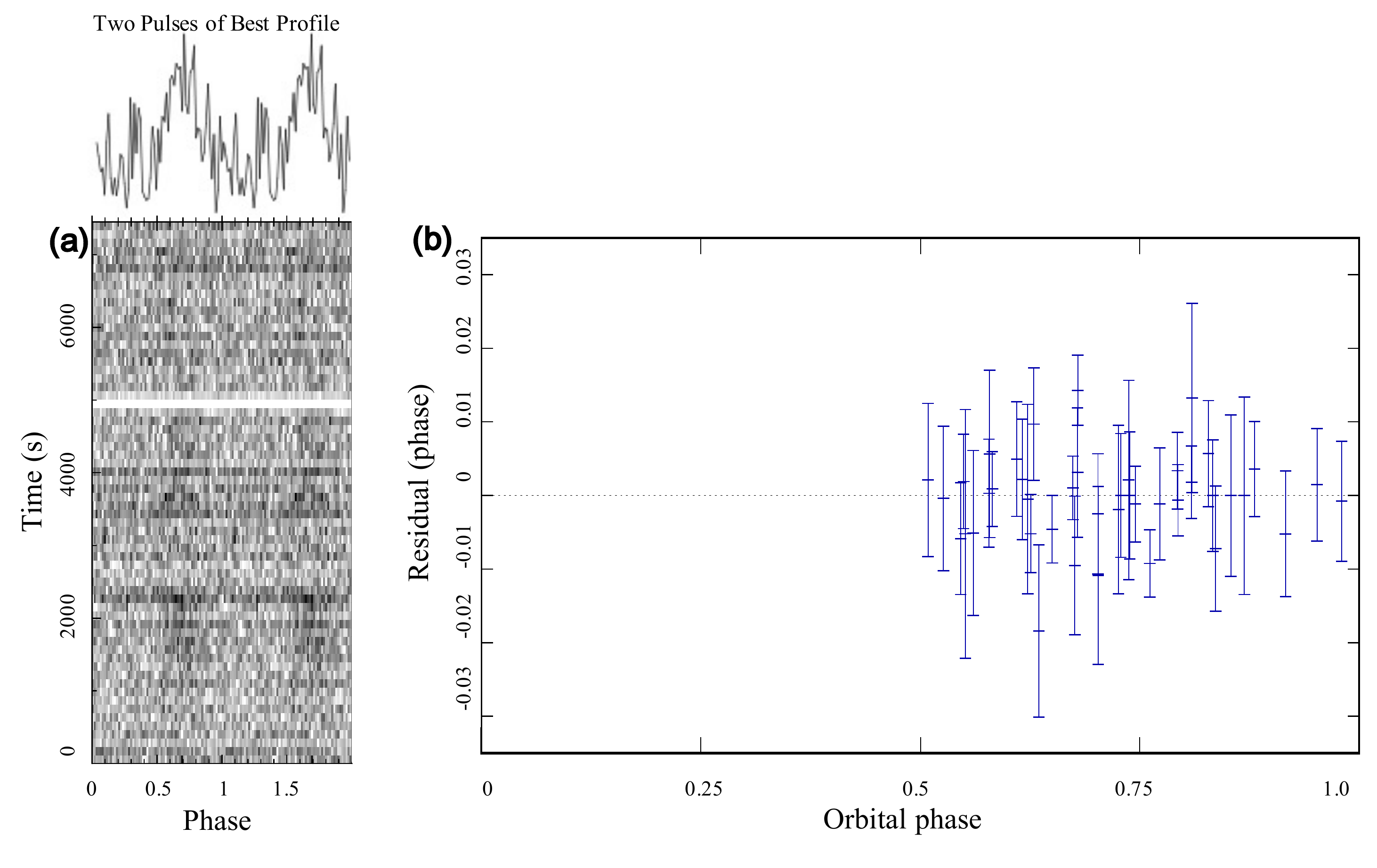}
\end{center}
 \caption{{\footnotesize{(a) The discovery diagnostic plot of MSP NGC~6397B, showing the result of 1.7\,h of folded search-mode data recorded by the Parkes UWL on April 12, 2019. The white horizontal stripe indicates a portion of data excised due to radio frequency interference. (b) Timing residuals versus orbital phase for NGC ~6397B. The pulsar is regularly eclipsed for 50\% of the orbit around its superior conjunction (i.e. around orbital phase 0.25), and it is therefore visible only when it is on the near side of the orbit (orbital phase range 0.5--1.0).}}}
\label{fig.NGC6397B}
\end{figure*}

\begin{table*}[]
\centering
\footnotesize
\caption{\footnotesize Radio Observations of pulsars in NGC 6397.}\label{tb.NGC6397_obs}
\setlength{\tabcolsep}{6mm}{
\begin{tabular}{cccccc}
\hline
\renewcommand{\arraystretch}{0.7}
No. & Start Date$^a$   & Start MJD$^b$   & Duration$^c$ & Pulsars & Telescope  \\
    &       & (MJD)    & (hours) & detected &          \\ \hline
1 & 2019 Apr 12  & 58585.543 & 10.9 &A, B & Parkes\\
2 & 2020 Apr 21  & 58960.780 & 4.4 &A, B & Parkes\\
3 & 2020 Apr 23  & 58962.760 & 4.7 &A, B & Parkes\\
4 & 2020 May 05  & 58974.782 & 3.4 &A, B & Parkes\\
5 & 2020 May 07  & 58976.770 & 1.0 &A, B   & Parkes \\
6 & 2020 May 17 & 58986.753 & 3.3 &A, B   & Parkes\\
7 & 2020 May 19 & 58988.720 & 3.9 &A, B & Parkes\\
8 & 2020 Jun 02 & 59002.723 & 3.0 &A, B & Parkes\\
9 & 2020 Jun 08 & 59008.718 & 2.9 &A, B & Parkes\\
10 & 2020 Jun 09 & 59009.633 & 3.2 &A & Parkes\\
11 & 2020 Jun 14 & 59014.392 & 9.2 &A, B & Parkes\\
12 & 2020 Jun 29 & 59029.346 & 8.8 &A, B & Parkes\\
13 & 2020 Jul 26 & 59056.546 & 1.3 &A, B & Parkes\\
14 & 2020 Oct 20 & 59142.624 & 3.3 &A, B & MeerKAT\\
15 & 2020 Nov 05 & 59158.425 & 4.0 &A, B & MeerKAT\\
16 & 2020 Dec 17 & 59200.151 & 1.0 &A & Parkes\\
17 & 2021 Mar 27 & 59300.982 & 1.3 &--  & Parkes\\
18 & 2021 May 04 & 59338.490 & 0.46 &A  & Parkes\\
19 & 2021 May 04 & 59338.512 & 10.1 &A  & Parkes\\
20 & 2021 May 31 & 59365.716 & 3.0 &A  & Parkes\\
21 & 2021 Jun 27 & 59392.466 & 2.7 &-- & Parkes\\
22 & 2021 Jul 26 & 59421.463 & 5.0 &A & Parkes\\
23 & 2021 Spe 26 & 59483.362 & 4.5 &A & Parkes\\
24 & 2021 Oct 03 & 59490.401 & 2.8 &A & Parkes\\
25 & 2021 Oct 31 & 59518.330 & 2.7 &A & Parkes\\
26 & 2021 Dec 18 & 59566.093 & 3.0 &A & Parkes\\
27 & 2022 Jan 08 & 59587.882 & 2.0 &A & Parkes\\
28 & 2022 Jan 13 & 59592.172 & 1.6 &A & Parkes\\
29 & 2022 Fer 17 & 59627.695 & 1.9 &A, B & Parkes\\
30 & 2022 Fer 26 & 59636.671 & 0.3 &A & Parkes\\
31 & 2022 Fer 27 & 59637.660 & 0.6 &A, B & Parkes\\
32 & 2022 Mar 10 & 59648.946 & 3.3 &A & Parkes\\
33 & 2022 Apr 03 & 59672.841 & 1.8 &-- & Parkes\\
34 & 2022 Apr 10 & 59679.550 & 4.5 &A & Parkes\\
35 & 2022 Apr 11 & 59680.879 & 1.0 &A, B & Parkes\\
36 & 2022 Apr 12 & 59681.857 & 1.4 &A & Parkes\\
37 & 2022 Apr 14 & 59683.856 & 1.4 &A & Parkes\\
38 & 2022 Apr 16 & 59685.858 & 1.4 &A & Parkes\\
39 & 2022 May 08 & 59707.723 & 2.0 &A & Parkes\\
\hline
\end{tabular}}
\begin{tablenotes}
\footnotesize
\item ~~~~~~~$^a$ The start date of the observation is considered in Coordinated Universal Time (UTC).
\item ~~~~~~~$^b$ Modified Julian date (MJD).
\item ~~~~~~~$^c$ Each observation length.
\end{tablenotes}
\end{table*}

\begin{table}[]
\centering
\caption{Parameters for NGC~6397B}\label{tb:NGC639B_par}
\setlength{\tabcolsep}{1mm}{
\begin{tabular}{lllll}
\hline
\multicolumn{3}{l}{Parameter}                    & \multicolumn{2}{l}{Value} \\ \hline
\multicolumn{5}{c}{Timing Parameters}                                        \\ \hline
\multicolumn{3}{l}{Right Ascension (J2000)$^{a}$}                 & \multicolumn{2}{l}{$17^{\rm h}\;40^{\rm m}\;42^{\rm s}.626$} \\
\multicolumn{3}{l}{Declination (J2000)$^{a}$}                & \multicolumn{2}{l}{$-53^{\circ}\;40^{'}\;27^{''}.91$}    \\
\multicolumn{3}{l}{Spin frequency, $\nu$ ($\rm s^{-1}$)} & \multicolumn{2}{l}{$172.802046855(3)$} \\
\multicolumn{3}{l}{Spin frequency derivative, $\dot{\nu}$ ($\rm s^{-2}$)} & \multicolumn{2}{l}{$1.77(95)\times 10^{-16}$} \\
\multicolumn{3}{l}{Timing epoch (MJD)} & \multicolumn{2}{l}{$59000$} \\
\multicolumn{3}{l}{Dispersion measure, DM (pc cm$^{-3}$)} & \multicolumn{2}{l}{$72.2$} \\
\multicolumn{3}{l}{Time units}                 & \multicolumn{2}{l}{TDB}\\
\multicolumn{3}{l}{Solar System ephemeris model} & \multicolumn{2}{l}{DE421} \\ 
\multicolumn{3}{l}{Binary model}                 & \multicolumn{2}{l}{BTX}    \\
\multicolumn{3}{l}{Orbital Frequency, $F_b$ (Hz)}   & \multicolumn{2}{l}{$5.853607(4)\times 10^{-6}$} \\
\multicolumn{3}{l}{Orbital Frequency Derivative$^{c}$, $\dot{F}_b$ (Hz$^{-2}$)}  & \multicolumn{2}{l}{$-1.4(5)\times 10^{-17}$} \\
\multicolumn{3}{l}{Projected semimajor axis, $\chi$ (ls)}    & \multicolumn{2}{l}{$2.58020(5)$} \\
\multicolumn{3}{l}{Time of ascending node, $T_{\rm asc}$ (MJD)}  & \multicolumn{2}{l}{ 58961.710672(11)} \\
\multicolumn{3}{l}{Total time span (days)$^{b}$}                 & \multicolumn{2}{l}{$1095$}    \\ 
\multicolumn{3}{l}{Number of ToAs}  & \multicolumn{2}{l}{$42$} \\
\multicolumn{3}{l}{Timing residual ($\mu s$)}  & \multicolumn{2}{l}{$33.90$} \\
\hline
\multicolumn{5}{c}{Derived Parameters}                                       \\ \hline
\multicolumn{3}{l}{Galactic longitude, $l$}          & \multicolumn{2}{l}{$338^{\circ}.165$}  \\ 
\multicolumn{3}{l}{Galactic longitude, $b$}          & \multicolumn{2}{l}{$-11^{\circ}.960$}  \\ 
\multicolumn{3}{l}{Minimum companion mass, $M_{c,min} (M_{\odot})^{d}$}   & \multicolumn{2}{l}{$0.2326$}  \\ 
\multicolumn{3}{l}{Median companion mass, $M_{c,med} (M_{\odot})^{d}$}   & \multicolumn{2}{l}{$0.2730$}  \\ 
\hline
\multicolumn{2}{l}{{\bf Note:} Numbers in parentheses represent uncertainties on the}\\
\multicolumn{2}{l}{last digit.}\\
\multicolumn{2}{l}{$^{a}$ Fixed at the optical position~\citep{Zhao20},  not fit}\\ 
\multicolumn{2}{l}{~~for in radio timing.}\\
\multicolumn{2}{l}{$^{b}$ Time between the first and 35th observations in Table~\ref{tb.NGC6397_obs}.}\\
\multicolumn{2}{l}{$^{c}$ For the orbital model, we also fitted eight more frequency } \\
\multicolumn{2}{l}{~~derivatives, the values of them are $1.0(3)\times10^{23}$, }\\
\multicolumn{2}{l}{~~$-2.7(8)\times10^{30}$, $-6.8(32)\times10^{38}$, $1.7(4)\times10^{43}$, $-1.9(6)\times10^{50}$,}\\
\multicolumn{2}{l}{~~$-3.7(9)\times10^{57}$, $9.9(26)\times10^{64}$,
$-6.4(17)\times10^{71}$.}\\
\multicolumn{2}{l}{$^{d}$ The companion masses assume a pulsar mass of 1.4$M_{\odot}$. The}\\
\multicolumn{2}{l}{~~minimum and median masses assume an inclination angle of}\\
\multicolumn{2}{l}{~~$90^{\circ}$ and $60^{\circ}$, respectively.}
\end{tabular}}
\end{table}

\section{Observations}
We carried out observations of NGC 6397 with the Parkes and MeerKAT radio telescope in the radio band, and with \textit{XMM-Newton} in the X-rays. In the following, we describe the observations made with each telescope.

\subsection{Parkes}
We have carried out 37 observations of NGC 6397 between 2019 April 12 and 2022 April 16 using the UWL receiver system on the Parkes 64-m radio telescope through different observing projects (P1006/PX500/PX501/P1128). The telescope was pointed at the nominal cluster center: $17^{\rm h}\;40^{\rm m}\;42^{\rm s}.09$, $-53^{\circ}\;40^{'}\;27^{''}.6$~\citep{Harris10}. Table~\ref{tb.NGC6397_obs} lists each observation.

Until 2021, the data were recorded with 2-bit sampling every 64\,${\mu}s$. For the observations from 2021 to now, the data were recorded with 8-bit sampling every 64\,${\mu}s$ in each of the 1\,MHz wide frequency channels along with Full Stokes information. A pulsed noise-diode signal injected into the signal path was observed before each observation to allow for calibration. In all observations,the observing band (from 704 to 4032\,MHz) was split into 3328, 1-MHz-wide frequency channels, which were also coherently de-dispersed at a DM of 71.80\,pc cm$^{-3}$ (corresponding to the first confirmed pulsar, NGC 6379A, in the GC).

\subsection{MeerKAT}
NGC~6397 was observed twice with MeerKAT, as part of the TRAPUM\footnote{\url{http://www.trapum.org}} (TRAnsients and PUlsars with MeerKAT, \citealt{Stappers16}) project. The two observations were acquired on 2020 October 20 and 2020 November 5. On both occasions, the cluster was observed for 4 hours using the L-band receivers, which cover the 856--1712 MHz frequency range. The FBFUSE computing cluster was used to syntesize 288 tied-array beams on the sky. The latter covered an area of about 2 arcmin in radius, centred around the nominal position of NGC~6397, with an hexagonal tiling pattern. The APSUSE computing cluster was used to split the observing band into 4096 channels, sample it every 76.56 $\mu$s and record the data of each tied-array beam as filterbank search-mode files.
These were subsequently incoherently de-dispersed using the dispersion measure (DM) of pulsar A, 71.8 pc\,cm$^{-3}$, and subbanded by summing the frequency channels in groups of 16. This significantly reduced the total data volume, making their analysis much faster. 
\subsection{XMM-Newton}
We obtained an \textit{XMM-Newton} observation of NGC 6397 on 31 August 2021. The European Photon Imaging Camera (EPIC) PN data were taken in timing mode while the MOS (Metal Oxide Semi-conductor; M1/M2) data were taken in Small-Window imaging mode. The total effective exposure times for MOS and PN cameras were 54 ks and 28 ks, respectively (i.e., a total of 82 ks).

\section{Combined Analyses and Results}

We searched for the hidden MSP in the first Parkes observations using the \textsc{presto} software suite~\citep{Ransom02}, with a Fourier drift-rate $z$ range~\citep{Andersen18} of $\pm 200$. This allowed us to be sensitive to short-period binary orbits pulsars~\citep{Ng15}. We searched within a DM range of 67--77\,pc cm$^{-3}$ using a relatively wide sub-band (960--3008\,MHz), which was cleaned of radio frequency interference (RFI).
We found a faint, 5.78-ms pulsar-like signal at a DM of 72.2 \,pc cm$^{-3}$ in multiple observations{, each time with a different associated acceleration, suggesting the presence of a binary motion.}  Being the second pulsar known in the cluster, the pulsar was dubbed PSR J1740-5340B (or NGC 6397B). 

To have an independent confirmation of the new pulsar, we also searched the innermost tied-array beams of the two MeerKAT observations, using \textsc{pulsar\_miner}\footnote{\url{https://github.com/alex88ridolfi/PULSAR_MINER}}~\citep{Ridolfi21}, a PRESTO-based automated pulsar searching pipeline. The signal was clearly detected in the central beam of both observations, thereby providing confirmation of the discovery.

 In order to characterize NGC 6397B, we performed a radio timing analysis as follows. First we folded each search-mode observation with 60\,s sub-integrations and 128 phase bins using \textsc{dspsr}\footnote{\url{http://dspsr.sourceforge.net}}~\citep{vanStraten11}. We removed data affected by radio frequency interference (RFI) in frequency and time for each channel and sub-integration, and then extracted pulse times of arrival (TOAs) from all Parkes detections. This was done using the software package \textsc{psrchive}\footnote{\url{http://psrchive.sourceforge.net}}~\citep{Hotan04} by averaging each observation in time, frequency and polarisation to obtain several high signal-to-noise profiles which were then correlated against a reference profile. The \textsc{tempo}\footnote{http://tempo.sourceforge.net} timing software  was then used to develop a comprehensive timing model of the pulsar's behaviour, including its spin frequency, spin frequency derivative, and a BTX binary model which assumed an eccentricity of zero. The position of the pulsar was not fitted, rather it was held fixed at the best determined optical position~\citep{Zhao20}. To solve for the unknown number of missing rotations between observations, and try to derive a unique phase-connected solution for the pulsar, we used the \textsc{dracula} code\footnote{\url{https://github.com/pfreire163/Dracula}}~\citep{Freire18}. However, a unique rotation count could not be determined for three large observing gaps (between MJDs 58585--58960, 59056--59627, and 59637--59680); for these gaps multiple numbers of rotations yield a good overall reduced $\chi^2$ of the residuals. For this reason, we cannot assume we know those rotation numbers, we must then instead fit for arbitrary time offsets between those data sets in order to derive timing parameters with conservative uncertainties. We note that the final model also includes a number of orbital frequency derivatives, a common requirement in the timing of redback systems. While these derivatives accurately describe the modeled timing span of the pulsar, they limit the timing model's predictive power outside of this span.
 
 For each observation where the pulsar was not detected, we carried out a brute-force search in orbital phase through the period–acceleration diagram based on this timing model using \textsc{spider\_twister}\footnote{\url{https://github.com/alex88ridolfi/SPIDER_TWISTER}}. No additional detections were obtained with this method.

Nevertheless, using the radio detections obtained from the first fifteen observations, we managed to successfully constrain the key orbital parameters of NGC~6397B. The pulsar is part of a binary system with an orbital period of 1.97\,days, consistent with that of the U18 source~\citep{Marcano21}. Timing parameters and uncertainties are given in Table~\ref{tb:NGC639B_par}. The optical position of U18~\citep{Zhao20} was a crucial input in finding the timing solution. Given the sensitive dependence on location when connecting the phase, the existence of the timing solution strongly supports NGC 6397B being U18. The pulsar signal is regularly eclipsed for about 50\% of the orbital phase, as evident in the timing residuals versus orbital phase plot (Figure~\ref{fig.NGC6397B}b). Typical of redbacks, in which matter lost by the companion absorbs the radio signals of the pulsar, NGC~6397B is detectable only when the pulsar is on the side of its orbit closer to the observer. Our confirmation of NGC 6397B being the hidden MSP candidate U18~\citep{Zhao20} also rules out the X ray source being a quiescent BH~\citep{Gallo14}.

\begin{figure}
\centering
\includegraphics[angle=0,width=0.5\textwidth]{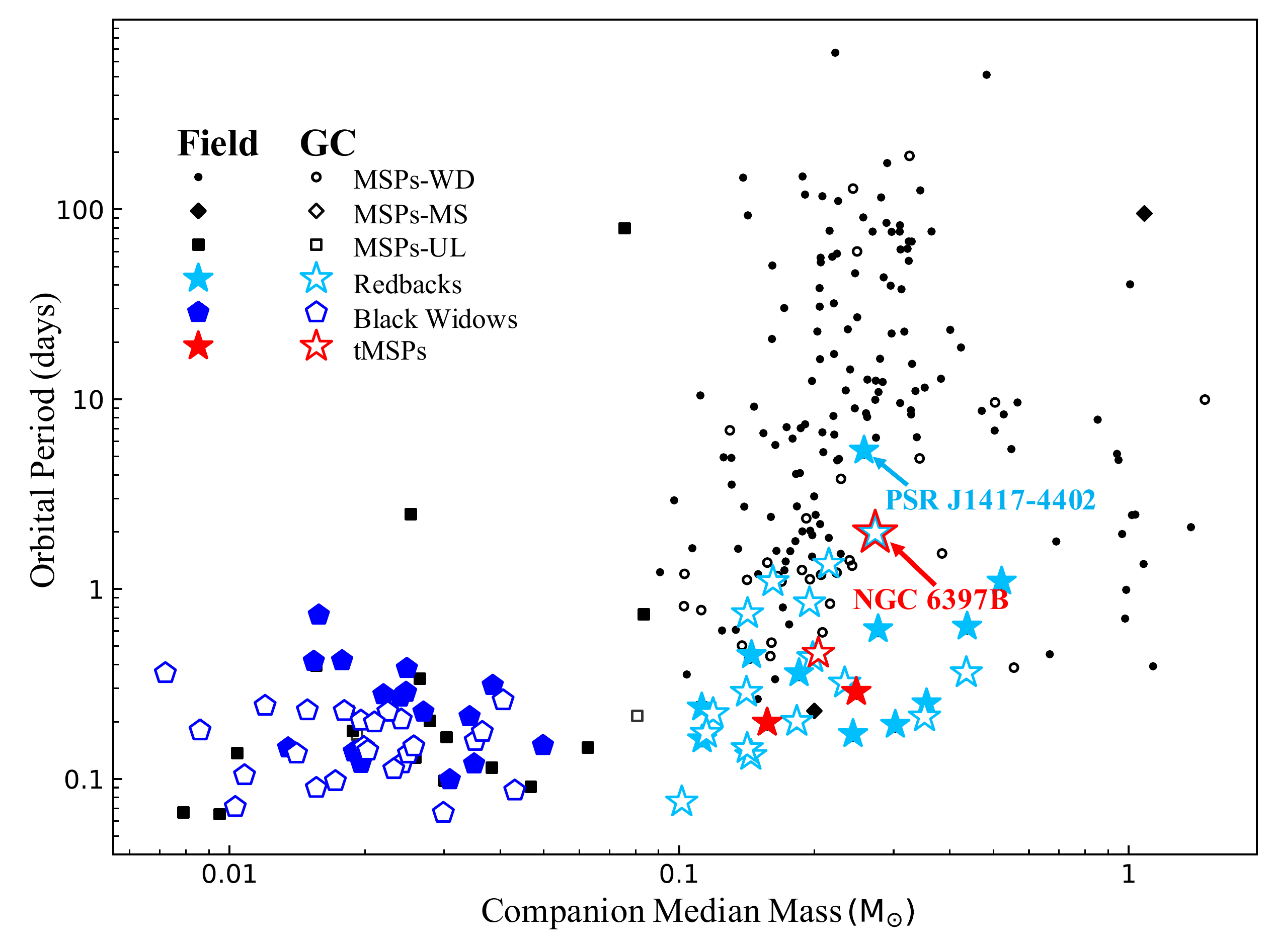}
\caption{\footnotesize Orbital period and median mass of radio MSPs (grey symbols with shape depending on the companion type, diamonds for WDs, circles for main-sequence stars and squares for Ultra-light companion or planet mass $<<0.08 M_{\odot}$), black widows (blue pentagons), redbacks (cyan stars) and transitional pulsars (tMSPs, magenta stars). Hollow and filled symbols mark sources found in GCs and the Galactic field, respectively.}
\label{fig.MSP_Mass-Pb}
 \end{figure}

 While the previously-known pulsar NGC~6397A was relatively easy to detect in nearly all of our observations, NGC~6397B was undetectable between  December 17, 2020 and January 13, 2022, a span of approximately 14\,months, despite being observed always around its inferior conjunction. Recently, NGC~6397B was re-detected at Parkes on February 17, 2022. This behaviour can be interpreted as resulting from scattering of the radio pulsations by a wind from the companion~\citep{Zhao20}.  The different detection rate for these two redbacks suggests that different eclipse mechanisms operate in the circum-pulsar environments of NGC~6397A and NGC~6397B~\citep{Zhao20} and/or significantly different wind parameters from the companion star. 

 Alternatively, the extended span of non-detections of NGC~6397B could imply that it is a transitional pulsar, which is a pulsar swinging between an accretion powered X-ray binary state and a rotation powered radio pulsar state, with known members of this class including PSRs~J1023+0038~\citep{Archibald09}, J1824$-$2452I~\cite{Papitto13}, and J1227$-$4853~\citep{Bassa14}. During a mass transfer phase, radio emission from these pulsars typically ceases, while an increase in X-ray emission is often detected.
 
 The \textit{XMM-Newton} observation was made during the radio-quiescent phase NGC 6397B, to check this hypothesis. We analyzed the \textit{XMM-Newton} data with the Science Analysis System (SAS\footnote{\url{https://www.cosmos.esa.int/web/xmm-newton/sas-threads}}, v12.0.1) while we conducted the spectral analysis with XSPEC\footnote{\url{https://heasarc.gsfc.nasa.gov/xanadu/xspec}}(v12.10.1f). We created light curves with 100 s bins, and the bins with count-rates greater than 0.1 and 0.2 counts/s were rejected for MOS and PN, respectively, thus filtering out spurious and heavy proton flaring events. We extracted the pulsar spectra from a 20$^{''}$ circular region from the MOS data, encompassing 77\% of the encircled energy, and the background spectra were extracted from an annular region of radius between 30$^{''}$ and 40$^{''}$ centered on the pulsar. The spectra were grouped to have a minimum of 15 counts per bin for the MOS data in the 0.5--10 keV band. We used the \textsc{tbabs} absorption model~\citep{Wilms00} to describe photoelectric absorption by the interstellar medium.  We first fitted the data with an absorbed blackbody (BB) model which did not provide an acceptable fit (reduced chi-squared $\chi^2$ = 437.98 for 179 degrees of freedom), with the high-energy end of the spectrum poorly characterized. Next, we fitted an absorbed powerlaw (PL) model, which gave the best-fit with a reduced chi-squared value $\chi^2$ = 152.26 for 179 degrees of freedom. The PL fit yielded the following results: $N_H$ = 1.0$^{+0.4}_{-0.4}$$\times$10$^{21}$~cm$^{-2}$, Photon index ($\Gamma$) = 1.2$^{+0.1}_{-0.1}$, and unabsorbed flux = 2$^{+0.3}_{-0.3}$$\times$10$^{-12}$ erg~cm$^{-2}$~s$^{-1}$ in the 0.5--10 keV energy range. Our analysis of the \textit{XMM-Newton} observation during the radio-quiet phase of NGC~6397B showed an unabsorbed flux of 2$^{+0.3}_{-0.3} \times 10^{-12} {\rm erg}~{\rm cm}^{-2}~{\rm s}^{-1}$, which is $\sim$20 times brighter than that ($\sim$1$\times$10$^{-13}$~erg~cm$^{-2}$~s$^{-1}$) obtained with the Chandra observations in 2007~\citep{Bogdanov10}. However, the 20$^{''}$ region around the pulsar encompasses four other sources near U18~\citep{Bogdanov10}, which were clearly resolved by Chandra. An independent analysis of the archival Chandra data including the four sources showed a flux similar to that obtained with our \textit{XMM-Newton} data, indicating no significant brightening as might be expected for the transitional state. Further optical observations to examine the presence (or lack of) of an accretion disk would be key for ascertaining the transitional status of NGC~6397B.

\section{Discussion}

 As of April 2022, we know of 660 MSPs (recycled pulsars with a spin period $<$ 30\,ms), 422 of which are in the Galactic field\footnote{\url{http://astro.phys.wvu.edu/GalacticMSPs/GalacticMSPs.txt}} and 238 are in 36 different GCs\footnote{\url{http://www.naic.edu/~pfreire/GCpsr.html}\label{GCpsr}}. MSPs in binaries can be further categorized according to the degeneracy of the companion star. Figure~\ref{fig.MSP_Mass-Pb} shows the 247 MSPs in binary systems, many of which are in compact interacting binaries, usually referred to as `spider’ MSPs. This class includes redbacks as well as `black widow' MSPs, which also experience eclipses but have much lower mass ($\leqslant 0.1 M_{\odot}$), degenerate companions~\citep{Roberts13}. When rotation powered, the three known transitional MSPs are classified as redbacks (see Figure~\ref{fig.MSP_Mass-Pb}). Notably, with a binary period of $\sim$1.97\,days, NGC~6397B has the longest orbital period among all known redback systems in GCs. Additionally, only a few redback systems in the Galactic field have orbital periods that exceed one day. Only PSR~J1417$-$4402, a redback system located in the Galactic field with known radio pulsations, has a longer orbital period than that of NGC 6397B at 5.3\,days~\citep{{Camilo16}}. With the long binary periods of NGC~6397B and PSR~J1417$-$4402, these two pulsars stand apart from the remaining black widows, redbacks, and other low-mass X-ray binaries that have binary periods $\leqslant 1$ day (see Figure~\ref{fig.MSP_Mass-Pb}), and suggests that NGC~6397B may be in the beginning stages of mass transfer after the companion star left the main sequence~\citep{Camilo16}.

 Globular clusters have also long been thought to harbor exotic compact objects, such as intermediate-mass black holes (IMBHs), dynamically disrupted systems, etc. NGC~6397 possesses a central dark component up to 2\% of its total mass~\citep{Vitral21}. As a core-collapsed cluster, NGC 6397 is unlikely to contain a large population of stellar-mass BH or a single IMBH at present~\citep{Rui21} and the dark component is deemed to primarily consist of massive WDs and NSs~\citep{Kremer21, Vitral22}. The identification of U18 to be NGC 6397B leaves the unidentified x-ray source U97~\citep{Zhao20} to be the only BH candidate in the cluster core. In the absence of BHs, NSs are among the most massive objects in NGC~6397 and thus are expected to efficiently mass-segregate to the cluster’s high density inner region. As a consequence, simulations of NS dynamics in GCs suggest that MSPs should form most readily in core-collapsed clusters~\citep{Ye19}. However, past surveys have confirmed only one pulsar in this cluster, despite the prediction  of roughly 300 NSs by N-body models of NGC~6397~\citep{Kremer21} and the prediction of about seven MSPs based on the 
 the correlation between the number of MSPs and stellar encounter rate~\citep{Zhao22}. The discovery of NGC~6397B  help reconcile the observed and expected populations of compact stars in this GC.

 Unlike in normal GCs, the vast majority of the pulsars discovered to date in core-collapsed GCs are isolated$^{\ref{GCpsr}}$. This is thought to be related to these clusters' very high encounter rate per binary, $\gamma$ \citep{Verbunt14}, a direct consequence of their exceedingly dense cores. Frequent encounters may have a tendency to break apart old, stable binary MSPs. However, simulations of such encounters suggest that, although many MSPs can become detached from their companions - with many even ejected from their GCs (as one can infer from the relatively numerous pulsars in core-collapsed clusters located far from their centres) - most should end up in new binaries, generally even tighter and more massive than the initial system~\citep{Phinney92}. For this reason the dearth of MSP binaries in core-collapsed GCs is somewhat puzzling.
 
 The detection of NGC~6397B, an extremely faint pulsar in the nearest core-collapsed GC, suggests a plausible explanation: a large fraction of the MSPs in the new binaries formed in such encounters are not easily detectable in radio bands. This is because, like NGC~6397B, they are either embedded in clouds of plasma originating in their main sequence or giant companion stars or are actively accreting matter from these companions as X-ray binaries, a condition known to inhibit radio emission \citep{Archibald09,Papitto13}.
 
This is not a problem for two classes of MSPs: first, those that become isolated after exchange encounters (but remained bound to their GC), which are therefore more likely to be detected in radio surveys of these clusters; second, those that acquire relatively massive degenerate binary companions (in highly eccentric orbits) through the secondary interactions. In both cases, orbital eccentricity and radio detectability can be preserved by lack of binary mass transfer. The idea that such secondary exchanges are especially important in core-collapsed GCs is supported by the fact that such exotic systems, unlike any in the Galactic disk (NGC 1851A, D and E, NGC 6544A, NGC 6652A, NGC 6624G and M15C, see \citealt{Ridolfi21,Ridolfi22} and references therein) are only observed in high-$\gamma$ globular clusters, particularly core-collapsed clusters. The small number of such systems indicates that their formation is relatively unlikely - in principle, acquiring a MS companion should be much more likely than acquiring a massive degenerate companion - however, the latter systems represent $\sim\,$1/3 of the binary pulsar population of core-collapsed GCs. This is an independent, but still rather qualitative indication, that many MSPs with MS companions in core-collapsed GCs may be missing in radio surveys.
 
As discussed by \cite{Verbunt14}, while the total interaction rate, $\Gamma$, yields a good prediction of the size of the LMXB and MSP populations in a GC, the interaction rate per binary $\gamma$ does explain qualitatively the most important differences between the pulsar populations of different GCs. Despite this, an assessment of the impact of $\gamma$ on the general properties of the pulsar (and other stellar) populations with detailed dynamical simulations has not been done to date, at least to our knowledge. If such a model is developed, it will be important to see whether the model does predict a smaller number of radio detections of binary MSPs (and thus MSPs in general) in core-collapsed clusters, and what the causes for this are.

\section{Conclusion}
With about two years of monitoring with the Parkes radio telescope and the MeerKAT radio telescope, we confirmed NGC~6397B to be the `hidden' millisecond pulsar known as U18, which was originally identified through X-ray and optical observations by Bogdanov et al. (2010). The identification of U18 as a hidden redback MSP via detection of its radio pulsations is exciting and significant in the following ways:\\
\begin{enumerate}[1.]
\item NGC 6397B is only the second pulsar in a cluster with an expected NS population in hundreds~\citep{Kremer21}. 

\item NGC 6397B has the longest orbital period, 1.97 days, for globular cluster eclipsing binaries (redback and black-widow systems). It is either a transitional MSP, or more likely an eclipsing binary in its initial stage of mass transfer after the companion has left the main sequence. 

\item The existence of NGC~6397B, with its faintness in radio and extended radio-quiescent epochs, provides a plausible explanation to the apparent over-abundance of isolated pulsars in the collapsed cores of GCs, where the stellar interactions have been expected to preferably result in binaries. 

\item NGC 6397 is the closest core-collapsed globular cluster and possesses a central dark component comprising of 2\% of the cluster’s mass~\citep{Vitral21}. It is thus surprising for past surveys to confirm only one pulsar, along with a few unidentified x-ray sources which exhibit signs of being NSs or stellar-mass BHs. The discovery of NGC 6397B leaves the unidentified x-ray source U97 as the only remaining candidate BH in the cluster core, and thus provides a critical additional constraint on the evolution of this GC and its underlying compact object population.

\item The characteristics of NGC~6397B suggest a possible explanation for the general lack of binaries in core-collapsed globular clusters. Such an explanation should be tested by detailed dynamical models.

\end{enumerate}

\section*{Acknowledgments}
This work is supported by the National Nature Science Foundation of China (NSFC) under grant No. 11988101, 11725313, 12103069, U2031121, U1731238, U2031117, 12041305, 12173016, 12041301, 12121003.
  L.Z. is supported by ACAMAR Postdoctoral Fellowship and especially thanks Matthew Bailes and Ryan Shannon at Swinburne University of Technology for their constructive feedback on this work. 
  K.K. is supported by an NSF Astronomy and Astrophysics Postdoctoral Fellowship under award AST-2001751.
  Y.F. is supported by the National Key R$\&$D Program of China No. 2017YFA0402600, and by Key Research Project of Zhejiang Lab No. 2021PE0AC03. 
  X.F. and Z.Y.Z acknowledge the science research grants from the China Manned Space Project with NO.CMS-CSST-2021-A08, CMS-CSST-2021-A07, the Program for Innovative Talents Entrepreneur in Jiangsu, the China Postdoctoral Science Foundation No. 2020M670023, and the National Key R\&D Program of China No. 2019YFA0405500. 
  P.W. acknowledges support from the Youth Innovation Promotion Association CAS (id. 2021055), CAS Project for Young Scientists in Basic Reasearch (grant YSBR-006) and the Cultivation Project for FAST Scientific Payoff and Research Achievement of CAMS-CAS.
  Q.J.Z. is supported by the Guizhou Provincial Science and Technology Foundation (Nos. [2016]4008, [2017]5726-37, [2018]5769-02), the Foundation of Guizhou Provincial Education Department (No. KY (2020) 003).
  A.R., M.K, J.B., D.J.C., P.F., P.V.P., V.V.K and W.C. acknowledge continuing valuable support from the Max-Planck Society.
  A.R. and M.B. gratefully acknowledges financial support by the research grant ``iPeska'' (P.I. Andrea Possenti) funded under the INAF national call Prin-SKA/CTA approved with the Presidential Decree 70/2016.
  This work is partly supported by the Pilot Project for Integrated Innovation of Science, Education and Industry of Qilu University of Technology (Shandong Academy of Sciences) under Grant 2020KJC-ZD02, and the Aoshan Science and Technology Innovation Project under Grant 2018ASKJ01. 
  Parts of this research were conducted by the Australian Research Council Centre of Excellence for Gravitational Wave Discovery (OzGrav), through project number CE170100004. We also thanks Parkes team for their great effort to install and commission the UWL receiver system.

\end{document}